\documentclass[letterpaper, conference]{IEEEtran}

\usepackage[utf8]{inputenc}
\usepackage{amssymb}
\usepackage{amsmath}
\usepackage{color}
\usepackage{theorem}
\usepackage{cite}
\usepackage{url}
\usepackage{pgfplots}
\usetikzlibrary{plotmarks}
\usepackage[ruled,vlined,titlenumbered,linesnumbered]{algorithm2e}
\usepackage{algorithmic}

\newtheorem{theorem}{Theorem}
\newtheorem{definition}{Definition}
\newtheorem{lemma}{Lemma}

\newtheorem{problem}{Problem}

\newcommand{\Fqm}{\ensuremath{\mathbb F_{q^m}}}

\newcommand{\Fq}{\ensuremath{\mathbb F_{q}}}

\newcommand{\intervallincl}[2]{\ensuremath{[#1,#2]}}
\newcommand{\intervallexcl}[2]{\ensuremath{[#1,#2-1]}}

\newcommand{\Linpolyring}{\mathbb{L}_{q^m}\![x]}

\newcommand{\OCompl}[1]{\ensuremath{\mathcal{O}({#1})}}

\DeclareMathOperator{\defi}{def}
\newcommand{\defeq}{\overset{\defi}{=}}

\renewcommand{\bar}{\overline}

\DeclareMathOperator{\rk}{rk}

\DeclareMathOperator{\RRE}{RRE}

\newcommand{\LT}[1]{\textrm{LT}(#1)}

\renewcommand{\vec}[1]{\ensuremath{\mathbf{#1}}}
\newcommand{\Mat}[1]{\ensuremath{\mathbf{#1}}}
\newcommand{\vecelements}[1]{\ensuremath{(#1_0 \ #1_1 \ \dots \ #1_{n-1})}}

\newcommand{\Mooremat}[2]{\Mat{V}\!_{#1}( #2 )}

\newcommand{\qVan}{\emph{q-Vandermonde} }

\newcommand{\MoormatExplicit}[3]{
\begin{pmatrix}
#1_{0} & #1_{1} & \dots& #1_{#3-1}\\
#1_{0}^{[1]} & #1_{1}^{[1]} & \dots& #1_{#3-1}^{[1]}\\[-4pt]
\vdots &\vdots&\ddots& \vdots\\[-2pt]
#1_{0}^{[#2-1]} & #1_{1}^{[#2-1]} & \dots& #1_{#3-1}^{[#2-1]}\\
\end{pmatrix}}

\newcommand{\Q}{\mathbf Q}

\newcommand{\mycode}[1]{\ensuremath{\mathcal{#1}}}

\newcommand{\Subspacedist}[1]{d_s(#1)}

\newcommand{\myspace}[1]{\mathcal{#1}}
\newcommand{\Rowspace}[1]{\myspace{R}_q\!\left(#1\right)}

\newcommand{\Projspace}{\myspace{P}_q(n)}

\newcommand{\nTransmit}{\ensuremath{n_t}}
\newcommand{\nReceive}{\ensuremath{n_r}}
\newcommand{\insertions}{\ensuremath{\gamma}}
\newcommand{\deletions}{\ensuremath{\delta}}
\newcommand{\intPoly}{\ensuremath{Q\left(x,y_1,\dots,y_s\right)}}
\newcommand{\kwDeg}{\ensuremath{(1,k-1,\dots,k-1)}}

\IEEEoverridecommandlockouts

\begin{document}

\title{Efficient Interpolation-Based Decoding of Interleaved Subspace and Gabidulin Codes}

\author{\IEEEauthorblockN{Hannes Bartz$^{1}$, Antonia Wachter-Zeh$^{2}$}
\IEEEauthorblockA{$^{1}$Institute for Communications Engineering, Technische Universit\"at M\"unchen, Munich, Germany\\
$^{2}$Computer Science Department, Technion---Israel Institute of Technology, Haifa, Israel\\
\texttt{hannes.bartz@tum.de, antonia@cs.technion.ac.il}\thanks{H. Bartz's work is supported by the German Ministry of Education and Research in the framework of the Alexander von
Humboldt-Professorship and by the TUM Graduate School. A. Wachter-Zeh's work has been supported by a Minerva Postdoctoral Research Fellowship. The authors would like to thank Gerhard Kramer for helpful comments and fruitful discussions.
}
}}
\maketitle

\begin{abstract}
An interpolation-based decoding scheme for interleaved subspace codes is presented. 
The scheme can be used as a (not necessarily polynomial-time) list decoder as well as a probabilistic unique decoder.
Both interpretations allow to decode interleaved subspace codes beyond half the minimum subspace distance.
Further, an efficient interpolation procedure for the required linearized multivariate polynomials is presented and a computationally- and memory-efficient root-finding algorithm for the probabilistic unique decoder is proposed.
These two efficient algorithms can also be applied for accelerating the decoding of interleaved Gabidulin codes. 
\end{abstract}


\section{Introduction}\label{sec:introduction}
Subspace codes have been proposed as a tool for non-coherent networks, i.e., the network topology and the linear combinations performed by the intermediate nodes are not known by the transmitter and receiver \cite{koetter_kschischang, silva_rank_metric_approach}.
Several code constructions, upper bounds on the size, and properties of such codes were thoroughly investigated in~\cite{koetter_kschischang,KohnertKurz-LargeConstantDimensionCodes-2008,Etzion2009ErrorCorrecting,silva_rank_metric_approach,Xia2009Johnson,Skachek2010Recursive,Gadouleau2010ConstantRank,TrautmannManganielloRosenthal-OrbitCodes-2010,Etzion2011ErrorCorrecting,Bachoc2012Bounds}.
Subspace codes with efficient decoding algorithms include the \emph{Reed--Solomon} (RS)-like code construction by K\"otter and Kschischang \cite{koetter_kschischang} (KK codes) and the lifted \emph{maximum rank distance} (MRD) codes (see also \cite{Delsarte_1978,Gabidulin_TheoryOfCodes_1985,Roth_RankCodes_1991}) by Silva, K\"otter and Kschischang \cite{silva_rank_metric_approach}.
For decoding lifted MRD codes, the linear operations performed by the network must be reversed before starting the decoding process.
This step is called \emph{reduction} and allows to apply known error-erasure decoding schemes for MRD codes to lifted MRD codes, but with the additional computational cost of the reduction, see \cite{silva_rank_metric_approach}. 

An \emph{interleaved} MRD code consists of $s$ parallel matrices, which are codewords of $s$ MRD codes (in particular Gabidulin codes). 
One benefit of using lifted \emph{interleaved} MRD codes in network coding is a reduced overhead for large interleaving orders. 
In \cite{Loidreau_Overbeck_Interleaved_2006,SidBoss_InterlGabCodes_ISIT2010,WachterzehZeh-InterpolationInterleavedGabidulin} it was shown that probabilistic unique decoding as well as (not necessarily polynomial-time) list decoding of interleaved Gabidulin codes beyond half the minimum rank distance is possible.
The main problem of list decoding subspace and MRD codes is that the list size might be exponential in the code length, see~\cite{Wachterzeh_BoundsListDecodingRankMetric_IEEE-IT_2013}.
The approaches of Mahdavifar and Vardy \cite{Mahdavifar2010Algebraic, Mahdavifar2012Listdecoding} and Guruswami and Xing~\cite{GuruswamiXing-ListDecodingRSAGGabidulinSubcodes_2012} provide list decoding schemes for subcodes and modifications of KK and MRD codes.
Further, Trautmann, Silberstein and Rosenthal presented an approach for list decoding lifted Gabidulin codes \cite{TrautmannSilbersteinRosenthal-ListDecodingLiftedGabidulinCodes}; here the complexity grows exponentially in the dimension of the subspace.

In this paper, we present an interpolation-based decoding scheme for $s$-\emph{interleaved KK codes} without the need for {reduction} at the receiver.
This approach can be applied as a (not necessarily polynomial-time) list decoder or as a probabilistic unique decoder.
The main contribution in this paper is that we show how the desired multivariate polynomials can be constructed efficiently using an interpretation of the general linearized polynomial interpolation from \cite{Xie2011General}.
The computational complexity of the interpolation step is therefore reduced from $\OCompl{s\nReceive^3}$ to $\OCompl{s^2\nReceive(\nReceive-\tau)}$ operations in $\Fqm$, where $\nTransmit$ and $\nReceive$ are the dimensions of the transmitted and received spaces, respectively, and $\tau$ denotes the decoding radius.
Further, we propose a computationally- and memory-efficient root-finding algorithm for the unique decoder which reconstructs the message polynomials in $\OCompl{s^2k^2}$ operations in $\Fqm$, where $k$ is the number of data symbols from $\Fqm$.
This reduces the computational complexity for the root-finding step compared to \OCompl{s^3k^2} for the recursive Gaussian elimination from~\cite{WachterzehZeh-InterpolationInterleavedGabidulin}.
Both proposed algorithms can also efficiently solve the interpolation problem and root-finding system for the interleaved Gabidulin codes in \cite{WachterzehZeh-InterpolationInterleavedGabidulin}.

This paper is structured as follows. 
In Section~\ref{sec:preliminaries}, we give basic definitions and describe notation.
Section~\ref{sec:principle} explains the principle of our interpolation-based decoding algorithm, including calculating the maximum number of tolerated insertions and deletions, and clarifying how we generalize and improve principles from~\cite{koetter_kschischang}.
In Section~\ref{sec:listunique}, we outline how our ideas apply to list and unique decoding. 
Sections~\ref{sec:efficient} and \ref{sec:efficientRootFinding} provide efficient interpolation and root-finding algorithms, and Section~\ref{sec:entire_eff_decoding} describes the entire decoding procedure.
Finally, Section~\ref{sec:conclusion} concludes this paper.

\section{Preliminaries}\label{sec:preliminaries}
\subsection{Finite Fields and Subspaces}
Let $q$ be a power of a prime, and denote by $\Fq$ the finite field of order $q$ and by $\Fqm$ its extension field of degree $m$. 
$\Fq^n$ denotes the vector space of dimension $n$ over $\Fq$ and the projective space $\Projspace$ is the set of all subspaces of $\Fq^n$.

Matrices and vectors are denoted by bold uppercase and lowercase letters, respectively, such as $\vec{A}$ and $\Mat{a}$. 
The row space and rank over $\Fq$ of a matrix $\Mat{A} \in \Fq^{m \times n}$ is denoted by $\Rowspace{\Mat{A}}$ and $\rk(\Mat{A})$ and the kernel of $\Mat{A}$ is denoted by $\ker(\Mat{A})$.
We index vectors and matrices beginning from zero, and $\intervallincl{1}{n}$ denotes the set $\{1,2,\dots,n\}$.

For two subspaces $\myspace{U},\myspace{V}$ in $\Projspace$, let $\myspace{U}\oplus\myspace{V}$ be the smallest subspace containing the union of $\myspace{U}$ and $\myspace{V}$. 
The \emph{subspace distance} between $\myspace{U},\myspace{V}$ in $\Projspace$ is
\begin{equation}\label{eq:subspaceDistance}
	\Subspacedist{\myspace{U},\myspace{V}}=\dim(\myspace{U}\oplus\myspace{V})-\dim(\myspace{U}\cap \myspace{V}). 
\end{equation}

A \emph{subspace code} is a non-empty subset of $\Projspace$, and has minimum subspace distance $d_s$ when all subspaces in the code have distance at least $d_s$ and there is one pair of subspaces with distance exactly $d_s$.

As channel model we use the operator channel from \cite{koetter_kschischang}.
Such a channel has input and output alphabet $\Projspace$. 
The output $\myspace{U}$ is related to the input $\myspace{V}$ of $\dim(\myspace{V})=\nTransmit$ by
\begin{equation}
 \myspace{U}=\mathcal{H}_{\nTransmit-\deletions}(\myspace{V})\oplus \myspace{E}
\end{equation}
where $\mathcal{H}_{\nTransmit-\deletions}(\myspace{V})$ returns a random $(\nTransmit-\deletions)$-dimensional subspace of $\myspace{V}$, and $\myspace{E}$ denotes an error space of dimension~$\insertions$ with $\myspace{V}\cap\myspace{E}=\emptyset$.
The distribution of $\mathcal{H}_{\nTransmit-\deletions}(\myspace{V})$ is not important for the performance of the code and can be chosen to be uniform (see \cite{koetter_kschischang}).
The dimension of the received subspace $\myspace{U}$ is thus $\nReceive=\nTransmit-\deletions+\insertions$ and we call $\deletions$ the number of \emph{deletions} and $\insertions$ the number of \emph{insertions}.

\subsection{Linearized Polynomials}
Let $a^{[i]}\defeq a^{q^{i}}$ be the $q$-power of an element $a\in\Fqm$ for any integer $i$.
A nonzero polynomial of the form
$p(x)=\sum_{i=0}^{d}p_ix^{[i]}$
with $p_i\in \Fqm$, $p_d\neq 0$, is called a \emph{linearized polynomial} of $q$-degree $\deg_q(p(x))=d$, see \cite{Ore_OnASpecialClassOfPolynomials_1933,Lidl-Niederreiter:FF1996}.
For all $a,b\in\Fq$ and $x_1,x_2\in\Fqm$, we have $p(ax_1+bx_2)=ap(x_1)+bp(x_2)$. 
Given two linearized polynomials $p^{(1)}(x)$ and $p^{(2)}(x)$ of $q$-degree $d_1$ and $d_2$, their non-commutative composition $p^{(1)}(x)\otimes p^{(2)}(x)=p^{(1)}(p^{(2)}(x))$ is a linearized polynomial of $q$-degree $d_1+d_2$.
The set of all linearized polynomials with coefficients from $\Fqm$ forms a non-commutative ring $\Linpolyring$ with identity under addition ``+'' and composition ``$\otimes$''. 

We define the $\kwDeg$-weighted degree $\deg_w$ of a \emph{multivariate linearized polynomial} of the form $\intPoly=Q_0(x)+Q_1(y_1)+\dots+Q_s(y_s)$ as
\begin{align*}
 	\deg_w(Q(x,y_1,&\dots,y_s))=\max\big\{\deg_q(Q_0(x)), \\ 
	k-1&+\deg_q(Q_1(y_1)),\dots,k-1+\deg_q(Q_s(y_s))\big\}.
\end{align*}
For brevity we also denote the $s+1$-variate linearized polynomial $Q(x,y_1,\dots,y_s)$ by $Q$.
The total order $\prec$ on monomials is defined as 
\begin{equation*}
x^{[\ell+k-1]} \prec y_1^{[\ell]} \prec y_2^{[\ell]} \prec \dots \prec y_s^{[\ell]} \prec x^{[\ell+k]}. 
\end{equation*}
We identify uniquely the \emph{leading term} $\LT{Q}$ of any multivariate polynomial $\intPoly$ as the maximum normalized monomial under $\prec$.

The \qVan matrix of the vector $\vec{a}=\vecelements{a}\in \Fqm^n$ is defined as\\ 
\vspace{-1ex}
\begin{equation}
  \Mooremat{r}{\vec{a}}=\MoormatExplicit{a}{r}{n}.
\end{equation}
The rank of $\Mooremat{r}{\vec{a}}$ is $\min\{r,n\}$ if the elements $a_0,\dots a_{n-1}$ are linearly independent over \Fq, see \cite{Lidl-Niederreiter:FF1996}.

\subsection{Lifted Rank-Metric Codes}
For any fixed basis of $\Fqm$ over $\Fq$,
there is a bijective mapping between any vector $\vec{a} \in \Fqm^n$ and a matrix $\Mat{A} \in \Fq^{m \times n}$. 
We often switch between these two representations.

The minimum rank distance $d$ of a code $\mycode{C}\subseteq \Fqm^n$ is defined as
\vspace{-3pt}
\begin{equation}
d = \min_{\vec{x},\vec{y} \in \mycode{C}} d_r(\vec{x}, \vec{y}) \defeq \min_{\vec{x},\vec{y} \in \mycode{C}} \rk(\Mat{X}-\Mat{Y})
\vspace{-2pt}
\end{equation}
where $\Mat{X},\Mat{Y}$ are the matrix representations of $\vec{x},\vec{y}\in\mycode{C}$.
The Singleton-like bound on the minimum rank distance states that $d_r\leq n-k+1$ when $m \geq n$, see \cite{Delsarte_1978,Gabidulin_TheoryOfCodes_1985,Roth_RankCodes_1991}.
Codes which attain this bound are called MRD codes.
A special class of MRD codes are \emph{Gabidulin codes} \cite{Delsarte_1978,Gabidulin_TheoryOfCodes_1985,Roth_RankCodes_1991}, which are the analogs of Reed--Solomon codes in rank metric.
\emph{Interleaved Gabidulin codes} are a horizontal or vertical concatenation of $s$ Gabidulin codes, for details see \cite{Loidreau_Overbeck_Interleaved_2006,SidBoss_InterlGabCodes_ISIT2010,WachterzehZeh-InterpolationInterleavedGabidulin}.

Lifting an (interleaved) MRD code means that we append an identity matrix to each (transposed) code matrix and consider the row space of these composed matrices as a subspace code.
The subspace distance of this constant-dimension code is twice the rank distance of the (interleaved) MRD code~\cite{silva_rank_metric_approach}.
The linear combinations performed by the intermediate nodes in a network coding scenario are therefore tracked by the leading identity matrix and can be inverted by computing the reduced row echelon form of the received matrix, called a \emph{reduction}.

\subsection{Interleaved Subspace Codes}
The main difference of KK codes \cite{koetter_kschischang} to the lifting approach from \cite{silva_rank_metric_approach} is that the evaluation points of the code, namely $\alpha_0,\dots,\alpha_{n-1}\in\Fqm$ (see also Definition~\ref{def:interleaved_subspace}), are appended instead of the identity matrix.
For KK codes, the \emph{reduction} is not necessary, which reduces the complexity at the receiver side, see \cite{koetter_kschischang}. 
Motivated by lifted interleaved Gabidulin codes and the construction of K\"otter and Kschischang, we define \emph{interleaved subspace (KK) codes} as follows.
\vspace{-1pt}
\begin{definition}[Interleaved Subspace (KK) Code]\label{def:interleaved_subspace}
 Let $\mathcal{A}\!=\!\{\alpha_0,\dots,\alpha_{\nTransmit-1}\} \!\subset\! \Fqm$ with $\nTransmit\leq m$ be a set of linearly independent elements over $\Fq$ and let 
 \begin{align*}
 W_s\!&= \Rowspace{\mathcal{A}} \oplus \Fqm\oplus \dots \oplus \Fqm \\
	&= \{(\alpha,\beta^{(1)}\!,\dots, \beta^{(s)}):\alpha \in \Rowspace{\mathcal{A}}\!, \beta^{(1)}\!,\dots,\beta^{(s)}\!\! \in \Fqm\}.
 \end{align*}
 For fixed integers $k^{(1)},\dots,k^{(s)} < \nTransmit$, an interleaved subspace code of dimension $\nTransmit$ and interleaving order $s$ is defined as
 \begin{equation*}
	\Rowspace{\left\{ \big(\alpha_i, f^{(1)}(\alpha_i), \dots,f^{(s)}(\alpha_i)\big):i\in\intervallexcl{0}{\nTransmit}\right\}}
 \end{equation*}
 where $f^{(j)}(x)\in\Linpolyring$, $\deg_q(f^{(j)}(x))<k^{(j)}$,  $\forall j\in\intervallincl{1}{s}$.
\end{definition}
For $s=1$ this definition is equivalent to KK codes \cite{koetter_kschischang} and the code rate of this construction is $R=\frac{skm}{\nTransmit(\nTransmit+sm)}$.
\begin{lemma}[Minimum Distance]
 The minimum subspace distance of an interleaved subspace code as in Definition~\ref{def:interleaved_subspace} is
 \begin{equation*}
  d_{s,min}= 2\big(\nTransmit-\max_{j\in\intervallincl{1}{s}}\{k^{(j)}\}+1\big).
 \end{equation*}
\end{lemma}

\begin{IEEEproof}
 Let $\myspace{V}$ and $\myspace{V}'$ be two codewords generated by $f^{(1)}(x),\dots,f^{(s)}(x)$ and $g^{(1)}(x),\dots,g^{(s)}(x)$ with $q$-degrees less than $k^{(1)}\!,\dots,k^{(s)}$.
 Since $\dim(\myspace{V})=\dim(\myspace{V}')=\nTransmit$, the minimum distance is achieved by the minimum dimension of the intersection space $\myspace{V}\cap\myspace{V}'$.
 The dimension of $\myspace{V}\cap\myspace{V}'$ is minimal when the evaluation polynomials of maximum $q$-degree, say $f^{(j)}(x)$ and $g^{(j)}(x)$, are distinct and all other evaluation polynomials are identical.
 Suppose $\dim(\myspace{V}\cap\myspace{V}')=r$, i.e., $f^{(j)}(x)$ and $g^{(j)}(x)$ agree on $r$ linearly independent points.
 Since $\deg_q(f^{(j)}(x)),\deg_q(g^{(j)}(x)) = k^{(j)}-1$, it follows that $r \leq \max\{k^{(j)}\}-1$.
 Hence we have 
 \begin{align*}
  d_{s,min}&= \dim(\myspace{V}) + \dim(\myspace{V}') - 2\dim(\myspace{V}\cap\myspace{V}') \\
		&\geq 2(\nTransmit-\max\{k^{(j)}\}+1)
 \end{align*}
 and it is easy to show that there are always two codewords such that equality holds.
\end{IEEEproof}

Throughout this paper, we consider only $k^{(j)} = k$, $\forall j \in \intervallincl{1}{s}$, but our approach also holds for arbitrary $k^{(j)}$.

\section{Interpolation-Based Decoding}\label{sec:principle}
Our decoding approach is 
closely related to decoding interleaved Gabidulin codes \cite{WachterzehZeh-InterpolationInterleavedGabidulin} and
generalizes the decoding approach in \cite{koetter_kschischang}.
The decoding principle consists of two steps: an interpolation step and a root-finding step.

\subsection{Interpolation Step}\label{subsec:interpolation}
Let $(x_i, r_i^{(1)}, \dots, r_i^{(s)})$ for $i\in\intervallexcl{0}{\nReceive}$ denote a basis of the received subspace $\mathcal{U}$ and let the matrix $[\vec{x}^T, \vec{r}^{(1)T}, \dots, \vec{r}^{(s)T}] \in \Fqm^{\nReceive \times (s+1)}$ contain this basis as rows. 
For the interpolation step, we must solve the following interpolation problem.

\begin{problem}[Interpolation Problem]\label{prob:intProblem}
 Find a nonzero $(s+1)$-variate linearized polynomial of the form 
	\begin{equation}\label{eq:intPoly}
	\intPoly = Q_0(x) + Q_1(y_1) + \dots + Q_s(y_s),
	\end{equation}
which satisfies the following conditions for given integers $\nReceive,\tau,k$:
	\begin{itemize}
	\item[$\bullet$] $Q(x_i,r_i^{(1)},\dots,r_i^{(s)}) = 0$, $\forall i \in \intervallexcl{0}{\nReceive}$,
	\item[$\bullet$] $\deg_q(Q_0(x)) < \nReceive-\tau$,
	\item[$\bullet$] $\deg_q(Q_j(y_j))< \nReceive-\tau-(k-1)$, $\forall j \in \intervallincl{1}{s}$.
	\end{itemize}
\end{problem}
Denote the coefficients of~\eqref{eq:intPoly} by
%
$Q_0(x)=\sum_{j=0}^{\nReceive-\tau-1}q_{0,j}x^{[j]}$ and 
$Q_i(y_i)=\sum_{j=0}^{\nReceive-\tau-k}q_{i,j}y_i^{[j]}$.
%
We can find the coefficients of $\intPoly$ by solving a linear system of equations $\Mat{R}\cdot\vec{q}^T=\Mat{0}$ 
where \Mat{R} is an $\nReceive \times ((s+1)(\nReceive-\tau) -s(k-1))$ matrix:
\begin{align}\label{eq:intMatrix}
 \Mat{R}=\Big(\Mooremat{\nReceive-\tau}{\vec{x}}^T, & \,\Mooremat{\nReceive-\tau-k+1}{\vec{r}^{(1)}}^T, \dots \nonumber \\[-5pt] 
				&\hspace{7ex}\dots,\,\Mooremat{\nReceive-\tau-k+1}{\vec{r}^{(s)}}^T\Big),
\end{align}
and 
 $\vec{q}=(q_{0,0},\dots,q_{0,\nReceive-\tau-1}| \ 
		\dots \ |q_{s,0},\dots,q_{s,\nReceive-\tau-k})$.
\begin{lemma}\label{lem:nonzero_interpolpoly}
A non-zero polynomial $\intPoly$ fulfilling the above interpolation constraints exists if
 \begin{equation}\label{eq:decradius_het}
 \tau < \frac{s(\nReceive  - k +1)}{s+1}.
 \end{equation}
\end{lemma}

\begin{IEEEproof}
The number of linearly independent equations is $\nReceive$ which must be less than the number of unknowns 
in order to guarantee that there is a non-zero solution. 
Hence we have
   $\nReceive < \nReceive-\tau + s \left( \nReceive-\tau-k+1\right)$.
\end{IEEEproof}

The following theorem shows that all message polynomials are roots of $\intPoly$ under certain constraints.

\begin{theorem} 
Let $\intPoly \neq 0$ fulfill the interpolation constraints. 
If $\insertions\leq\tau$, where $\tau$ satisfies \eqref{eq:decradius_het}, then
\begin{equation}\label{eq:decodable}
P(x) \defeq Q\big(x,f^{(1)}(x),\dots,f^{(s)}(x)\big) = 0.
\end{equation}
\end{theorem}

\begin{IEEEproof}
 Let $(x_i, r_i^{(1)}, \dots, r_i^{(s)})$ for $i\in\intervallexcl{0}{\nTransmit-\deletions}$ denote the non-corrupted dimensions of the received subspace $\myspace{U}$, i.e., a basis for $\myspace{U} \cap \myspace{V}$.
 Due to the interpolation constraints, we have
 \begin{equation*}
  Q\big(x_i,f^{(1)}(x_i),\dots,f^{(s)}(x_i)\big)= 0, \ \forall i\in\intervallexcl{0}{\nTransmit-\deletions}
 \end{equation*}
 since $r_i^{(j)}=f^{(j)}(x_i)$ for $i\in\intervallexcl{0}{\nTransmit-\deletions}$ and $j\in\intervallincl{1}{s}$.
 The dimension of the root space of $P(x)$ over $\Fqm$ is at most $\nReceive-\tau-1$, since $\deg_q(P(x))<\nReceive-\tau$.
 If we choose
 \begin{equation}\label{eq:gammaleqtau}
  \tau \geq \insertions
  \quad\Longleftrightarrow\quad
  \nTransmit-\deletions \geq \nReceive-\tau=\nTransmit-\deletions+\insertions-\tau
 \end{equation}
 then the dimension of the root space of $P(x)$ is larger than its degree, since $(x_i, r_i^{(1)}, \dots, r_i^{(s)})$ for $i\in\intervallexcl{0}{\nTransmit-\deletions}$ are linearly independent. 
 This is possible only if $P(x)=Q\left(x,f^{(1)}(x),\dots,f^{(s)}(x)\right)= 0$.
\end{IEEEproof}
If we substitute \eqref{eq:decradius_het} into \eqref{eq:gammaleqtau}, we obtain the \emph{decoding radius}:
\begin{align}\label{eq:dec_condition}
 \insertions (s+1) &< s(\nReceive-k+1)=s(n_{t}-\deletions + \insertions-k + 1), \nonumber\\
	\Longleftrightarrow \quad \frac{\insertions}{s} + \deletions &< \nTransmit-k+1.
\end{align}
We say that the received subspace is \emph{decodable} if \eqref{eq:dec_condition} holds.
From \eqref{eq:dec_condition} we see that interleaving makes the code more resilient against insertions while the performance for decoding deletions remains the same as in \cite{koetter_kschischang}.
A similar behavior can be observed in \cite{Mahdavifar2012Listdecoding} for folded subspace codes.

\subsection{Root-Finding Step}\label{subsec:root-finding}
The task of the root-finding step is to find all polynomials $f^{(j)}(x)$ with $\deg_q(f^{(j)}(x))<k$, $\forall j\in\intervallincl{1}{s}$, such that \eqref{eq:decodable} holds.
Similar to \cite{WachterzehZeh-InterpolationInterleavedGabidulin}, this is done by solving a linear system of equations in $\Fqm$ for the coefficients of $f^{(1)}(x),\dots,f^{(s)}(x)$.

Instead of using only one solution of the interpolation step, we use a whole basis of the solution space for the root-finding step.
In order to determine the dimension of the solution space, we first derive the rank of the interpolation matrix.
\begin{lemma}[Rank of Interpolation Matrix]
 Let $\insertions\leq \tau$.
 The rank of the root-finding matrix satisfies 
 \begin{equation}\label{eq:rankRoot}
  \rk(\Mat{R})\leq \nTransmit-\deletions+2\insertions-\tau=\nReceive-\tau+\insertions.
 \end{equation}

\end{lemma}
\begin{IEEEproof}
 Assume w.l.o.g. that the first $\nTransmit-\deletions$ dimensions $(x_i,r_i^{(1)}\!,\dots,r_i^{(s)}), \forall i\!\in\!\intervallincl{0}{\nTransmit\!-\!\deletions\!-\!1}$, correspond to the basis of the non-corrupted subspace $\myspace{U}\cap \myspace{V}$.
 Since the elements $x_0,\dots,x_{\nTransmit-\deletions-1}$ are linearly independent over $\Fq$ and $\nReceive-\tau\leq \nTransmit-\deletions$, 
 the rank of the $(\nTransmit-\deletions) \times (\nReceive-\tau)$ $q$-Vandermonde matrix $\Mat{R}'=\Mooremat{\nReceive-\tau}{(x_0,\dots,x_{\nTransmit-\deletions-1})}$ equals
  $\rk\big(\Mat{R}')=\nReceive-\tau$.
 The rank of \Mat{R} depends on $\rk(\Mat{R}')$ and $\dim(\myspace{E})=\insertions$.
 Hence we have \eqref{eq:rankRoot}.
\end{IEEEproof}

The dimension of the solution space of the system of equations for the interpolation step thus satisfies
\begin{equation}\label{eq:dimsolutionstep}
 d_{I}\defeq\textrm{dim ker}(\Mat{R}) 
	\geq s(\nTransmit-k-\deletions-\tau+1)+(s-1)\insertions.
\end{equation}
We now set up the root-finding matrix using $d_I$ polynomials $Q^{(h)}(x,y_1,\dots,y_s)$, $h\in\intervallincl{1}{d_I}$,
whose coefficient vectors span the solution space of \eqref{eq:intMatrix}.
Define
\begin{equation*}
 	\Mat{Q}_{j}^{[i]} \defeq 
	\begin{pmatrix}
	q_{1,j}^{(1)[i]} & q_{2,j}^{(1)[i]} & \dots & q_{s,j}^{(1)[i]}\\[-3pt]
	\vdots &\vdots&\ddots& \vdots\\
	q_{1,j}^{(d_I)[i]} & q_{2,j}^{(d_I)[i]} & \dots & q_{s,j}^{(d_I)[i]}\\
	\end{pmatrix},
\end{equation*}
\begin{equation*}
 	\vec{f}_{j}^{[i]} \defeq 
	\left(f_{j}^{(1)[i]}\!\dots f_{j}^{(s)[i]}\right)^{\!T}\!\!
 	\text{ and  } 
	\vec{q}_{0,j}^{[i]} \defeq 
	\left(q_{0,j}^{(1)[i]}\!\dots q_{0,j}^{(d_I)[i]}\right)^{\!T}\!\!.
\end{equation*}
The root-finding matrix can be set up as
\begin{equation}\label{eq:rootFindingMatrix}
 \Mat{Q}=
 \begin{pmatrix}
 \Mat{Q}_0^{[0]}\phantom{^{-}}				&			&	 &							\\
 \Mat{Q}_1^{[-1]}					& \Mat{Q}_0^{[-1]}	&        &							\\[-3pt]
 \vdots  						& \Mat{Q}_1^{[-2]}	& \ddots &							\\[-3pt]
 \Mat{Q}_{\nReceive-\tau-k}^{[-(\nReceive-\tau-k)]}	& \vdots		& \ddots & \Mat{Q}_{0}^{[-(k-1)]}				\\[-6pt]
 							& \ddots		& \ddots & \vdots						\\[-5pt]
 							&      		 	& \ddots & \Mat{Q}_{\nReceive-\tau-k}^{[-(\nReceive-\tau-1)]}
 \end{pmatrix}
\end{equation}
and the roots can be found by solving the system of equations
\begin{equation}\label{eq:rootFindingSystem}
 \Mat{Q}\!\cdot\!
	\left(\vec{f}_0\,\vec{f}_1^{[-1]}\dots\vec{f}_{k-1}^{[-(k-1)]}\right)^T\!\!\!\!
	=
	\!\left(-\vec{q}_{0,0}\dots-\vec{q}^{[-(\nReceive-\tau-1)]}_{0,\nReceive-\tau-1}\right)^T\!.
\end{equation}
Solving the root-finding system \eqref{eq:rootFindingSystem} recursively requires at most $\OCompl{s^3k^2}$ operations in $\Fqm$ (see \cite{WachterzehZeh-InterpolationInterleavedGabidulin}).
\subsection{Connection to Decoding of K\"otter--Kschischang Codes}
We show in the following that the list-one decoding algorithm from~\cite{koetter_kschischang} is a special case of our proposed decoding scheme for $s=1$.
The decoding procedure in \cite{koetter_kschischang} interpolates a bivariate linearized polynomial $Q(x,y)=Q_0(x)+Q_1(y)$ and uses a root-finding step to extract the message polynomial $f(x)$ from $Q(x,f(x))=Q_0(x)+Q_1(x)\otimes f(x)$.

We compare the degree constraints for $Q(x,y)$ in \cite{koetter_kschischang} with our approach.
From~\eqref{eq:decradius_het} we obtain $\tau \!\leq\! \big\lceil \frac{s(\nReceive-k+1)-1}{s+1} \big\rceil$.
The degree of $Q_0(x)$ is thus less than
\begin{align*}
 \nReceive-\tau &= n_r\!-\!\left\lceil \frac{s(\nReceive-k+1)-1}{s+1} \right\rceil \!=\! \left\lceil \frac{\nReceive+s(k-1)+1}{s+1}\right\rceil\!.
\end{align*}
For $s=1$, this gives the degree constraint on $Q_0(x)$ from \cite{koetter_kschischang}. 
The same applies to the degree constraints on $Q_j(y_j), \forall j \!\in\! \intervallincl{1}{s}$ and 
decodability condition \cite[Eq. 11]{koetter_kschischang} is equivalent to~\eqref{eq:dec_condition} for $s=1$ and improves upon \cite{koetter_kschischang} for higher $s$.
\vspace{15pt}
%
\section{Application to List and Unique Decoding}\label{sec:listunique}

\subsection{List Decoding Approach}
In general, the root-finding matrix $\Mat{Q}$ \eqref{eq:rootFindingMatrix} does not always have full rank.
In this case, we obtain a \emph{list} of roots of \eqref{eq:decodable}, i.e., a list of possible (interleaved) message polynomials. 
This decoder is not a polynomial-time list decoder but it provides the basis of the list with quadratic complexity.
The derivation of the maximum and average list size is similar to \cite{WachterzehZeh-InterpolationInterleavedGabidulin}. 

\subsection{Probabilistic Unique Decoder}
The decoder can also be used as a probabilistic unique decoder.
We obtain a unique solution to the root-finding problem \eqref{eq:rootFindingSystem} if the rank of $\Mat{Q}$ is full.
If $\Mat{Q}$ does not have full rank, we declare a decoding failure.
We show in the following that the probability of a non-correctable error is very small. 

\begin{lemma}[Rank of Root-Finding Matrix]\label{lem:rootFinding}
 Let $\Mat{Q}$ be defined as in \eqref{eq:rootFindingMatrix}. 
 If $\rk(\Mat{Q}_{\nReceive-\tau-k})=s$ then $\rk(\Mat{Q})=sk$.
\end{lemma}

\begin{IEEEproof}
Note that $\rk(\Mat{Q}_{\nReceive-\tau-k}^{[i]})=\rk(\Mat{Q}_{\nReceive-\tau-k})$ for any integer $i$.
 Since $\Mat{Q}$ contains a lower block triangular matrix with $\Mat{Q}_{\nReceive-\tau-k}^{[-(\nReceive-\tau-k)]},\dots,\Mat{Q}_{\nReceive-\tau-k}^{[-(\nReceive-\tau-1)]}$,
 $\rk(\Mat{Q})=sk$ holds if $\rk(\Mat{Q}_{\nReceive-\tau-k})=s$.
\end{IEEEproof} 

Clearly, $\Mat{Q}_{\nReceive-\tau-k}$ can have rank $s$ only if $d_I\geq s$, which is guaranteed for $\insertions\leq\tau$ if
\begin{align}\label{eq:rankRootFinding}
	d_{I}\geq s \quad \Longleftrightarrow \quad \insertions &\leq\frac{s(\nReceive-k)}{(s+1)}= s(\nTransmit-k-\deletions).
\end{align}
We can derive an upper bound on the fraction of non-correctable errors similar to \cite{WachterzehZeh-InterpolationInterleavedGabidulin}. 
The proof is omitted due to space restrictions.

\begin{lemma}[Fraction of Non-Correctable Errors]\label{lem:failureproba_bound}
\hspace{5em}
Let $d_I \geq s$ (see \eqref{eq:dimsolutionstep}), let $\Mat{Q}$ be as in \eqref{eq:rootFindingMatrix}, and let $q^{(h)}_{1,\nReceive-\tau-k},\dots, q^{(h)}_{s,\nReceive-\tau-k}$ for $h\in\intervallincl{1}{d_I}$ be random elements uniformly distributed over $\Fqm$. 
Then the fraction of non-correctable errors is upper bounded by
%
\begin{equation*}
P\big(\rk(\Q) < sk\big) \leq 4q^{-(m(d_I+1-s))}. 
\end{equation*}
\end{lemma}
In a simulation with $10^6$ transmissions over an operator channel with $\delta=0$, $\gamma=5$ and code parameters $m=8$, $n_t=7$, $k=4$, $s=2$, we observed a fraction of $1.5\cdot 10^{-5}$ non-correctable errors (upper bound $6.1\cdot10^{-5}$). 
%
\section{Efficient Interpolation}\label{sec:efficient}
A single polynomial which is a solution to Problem~\ref{prob:intProblem} can be constructed efficiently by the general linearized polynomial interpolation algorithm by Xie, Yan and Suter \cite{Xie2011General}, requiring at most $\OCompl{s^2\nReceive(\nReceive-\tau)}$ operations in $\Fqm$. 
From Section \ref{subsec:root-finding} we know that probabilistic unique decoding is possible if we use a whole basis of the solution space of the interpolation problem for the root-finding step.
Hence we can reformulate the \emph{extended} interpolation problem as follows.
\begin{problem}[Extended Interpolation Problem]\label{prob:extInt}
Find $d_I\geq s$ nonzero $(s+1)$-variate linearized polynomials of the form 
\begin{equation*}\label{eq:extIntPoly}
 Q^{(h)}= Q^{(h)}_0(x) + Q^{(h)}_1(y_1) + \dots + Q^{(h)}_s(y_s)
\end{equation*}
which are linearly independent and satisfy the following conditions for given integers $\nReceive,\tau,k$ for all $h\in\intervallincl{1}{d_I}$:
\begin{itemize}
\item[$\bullet$] $Q^{(h)}(x_i,r_i^{(1)},\dots,r_i^{(s)}) = 0$, $\forall i \in \intervallexcl{0}{\nReceive}$,
\item[$\bullet$] $\deg_q(Q^{(h)}_0(x)) < \nReceive-\tau$,
\item[$\bullet$] $\deg_q(Q^{(h)}_j(y_j))< \nReceive-\tau-(k-1)$, $\forall j \in \intervallincl{1}{s}$
\end{itemize}
\end{problem}

Instead of constructing \emph{one} multivariate linearized polynomial, 
we therefore want to construct a \emph{set} of linearly independent polynomials that all vanish on a given set of points (and fulfill certain degree constraints).
This set of polynomials lies in the kernel of $\Mat{R}$ in \eqref{eq:intMatrix}.
Our main contribution in this section is to show how the general linearized polynomial interpolation algorithm \cite{Xie2011General} can be used to solve Problem~\ref{prob:extInt} efficiently.
We use notation from \cite{Xie2011General}.

Let $V=\{Q(y_0,y_1,\dots y_s)\}$ be a free (left) $\Linpolyring$-module constructed by the basis $\{y_0,y_1,\dots,y_s\}$.
Any element $Q\in V$ can be represented by $Q=\sum_{j=0}^{s}Q_j(x)\otimes y_j,$ where $Q_j(x)\in\Linpolyring$.
As in \cite{Xie2011General}, $V$ can be partitioned as $V=\bigcup_jS_j$, where $S_j=\{Q\in V:\LT{Q}=y_j^{[\ell]}\}$ for some $\ell\geq0$.
Hence, the subset $S_j$ contains all $Q\in V$ with leading term in $y_j$.
Further, define a set of $\nReceive$ functionals $D_i:V\mapsto\Fqm, i\in\intervallincl{1}{\nReceive}$:
\begin{equation*}
 D_i(Q)=Q\big(x_i,r_i^{(1)},\dots,r_i^{(s)}\big), \quad \forall i\in\intervallincl{1}{\nReceive}.
\end{equation*}
The kernel of $D_i$ is denoted by $K_i$ and we define $\bar{K}_i=K_1\cap K_2\cap\dots\cap K_i$.
Therefore, $\bar{K}_i$ contains all elements $Q\in V$ that are mapped to zero under $D_1,\dots,D_i$.

Algorithm 1 in~\cite{Xie2011General} iteratively constructs $s+1$ polynomials in each step $i$ that are minimal in the $\Linpolyring$-submodule $\bar{K}_i$ w.r.t. the weighted degree $\deg_w$.
Polynomials which have the leading term in $y_j$ of minimal degree are called $y_j$-minimal for $j\in\intervallincl{1}{s}$.
Let $T_{i,j}=\bar{K}_i\cap S_j$ contain all elements $Q\in V$ from the kernel with $\LT{Q}$ in $y_j$. 

The output of the algorithm is \emph{one} polynomial $Q^*$ which is minimal in $\bar{K}_{\nReceive}$.
Instead of using only the minimal $Q^*\in\bar{K}_{\nReceive}$, we use $g_{\nReceive,1},\dots,g_{\nReceive,s}$, which are minimal in $T_{\nReceive,1},\dots,T_{\nReceive,s}$. 
Hence we  modify the algorithm so that it outputs $g_{\nReceive,1},\dots,g_{\nReceive,s}$ and denote this adapted version by $\text{InterpolateBasis}(\vec{x}^{T}, \vec{r}^{(1)T}, \dots, \vec{r}^{(s)T})$.

From the root-finding step we know that there is a unique solution if $\rk(\Mat{Q}_{\nReceive-\tau-k})=s$. 
We now relate this condition to the interpolation algorithm and show that $s$ of the generated polynomials are a solution for the extended interpolation problem (Problem \ref{prob:extInt}).

%
\begin{theorem}\label{thm:minimalPolys}
If $\rk(\Mat{Q}_{\nReceive-\tau-k})=s$, then the linearized polynomials $g_{\nReceive,1},\dots,g_{\nReceive,s}$ constructed by $\text{InterpolateBasis}(\cdot)$ are linearly independent and have $(1,k-1,\dots,k-1)$-weighted degree less than $\nReceive-\tau$.
\end{theorem}
%
%
\begin{IEEEproof}
From \cite[Lemma 2]{Xie2011General} we know that after each step $i$, $g_{i+1,j}$ is a minimum element in $T_{i+1,j}, j\in\intervallincl{0}{s}$. 
This implies that the leading terms $\LT{g_{i+1,j}}$ are distinct.
Consequently, the $s+1$ polynomials are linearly independent. 
First we show, that $g_{\nReceive,0}$ is never a solution to the interpolation problem. Let $g_{\nReceive,0} = Q_0(x)+Q_1(y_1)+\dots+Q_s(y_s)$ be a non-zero polynomial fulfilling the interpolation constraints from Section \ref{subsec:interpolation}.
Since $g_{\nReceive,0}$ is the minimal in $T_{\nReceive,0}$ w.r.t. $\deg_w$, we have  
\begin{equation}
\label{eq:xMinimality}
 \deg_q(Q_0(x))> \max\{\deg_q(Q_j(y_j))+k-1\}.
\end{equation}
In order to fulfill $Q(x,f^{(1)}(x),\dots,f^{(s)}(x))= 0$ with $\deg_q(f^{(j)}(x))<k$ for $j\in\intervallincl{1}{s}$,
 $\deg_q(Q_0(x)) \leq \max\{\deg_q(Q_j(y_j))\!+\!k\!-\!1\}$
must hold, which contradicts \eqref{eq:xMinimality}.

Lemma~\ref{lem:rootFinding} shows that there is a unique solution to~\eqref{eq:rootFindingSystem} if $\rk(\Mat{Q}_{\nReceive-\tau-k})=s$.
Clearly, $\Mat{Q}_{\nReceive-\tau-k}$ contains the leading coefficients of $Q_{1}^{(h)}(y_1),\dots,Q_{s}^{(h)}(y_s)$ in each row $h\in\intervallincl{1}{d_I}$.
Suppose $\rk(\Mat{Q}_{\nReceive-\tau-k})=s$, then  the first $s$ rows of $\RRE(\Mat{Q}_{\nReceive-\tau-k})=I_{s\times s}$. 
Thus, there exist $s$ solutions to the interpolation problem $\Mat{R}\cdot\vec{q}^T=\vec{0}$ such that $q_{j,\nReceive-\tau-k}^{(h)}\neq 0$ and $q_{j,\nReceive-\tau-k}^{(h')}= 0$ for all $h\neq h'$ and $h,h',j\in\intervallincl{1}{s}$. 
Since $\deg_q(Q^{(h)}_0(x))<\nReceive-\tau$ is already implied by the structure of $\Mat{R}$, the $q_{j,\nReceive-\tau-k}^{(h)}$ are the leading coefficients of $Q^{(h)}(x,y_1,\dots,y_s)$.
Consequently, there exist $s$ polynomials with $\LT{Q^{(j)}(x,y_1,\dots,y_s)} = y_j^{[\ell]}$ for some $\ell$ that are contained in $T_{\nReceive,j}$ and $j\in\intervallincl{1}{s}$.
The polynomials $g_{\nReceive,1},\dots,g_{\nReceive,s}$ constructed by $\text{InterpolateBasis}(\cdot)$ are minimal in the corresponding $T_{\nReceive,j}$ and thus they are a solution to the interpolation problem if $rk(\Mat{Q}_{\nReceive-\tau-k})\geq s$ holds.
\end{IEEEproof}
Theorem \ref{thm:minimalPolys} implies that at least one of the $s$ polynomials $g_{\nReceive,1},\dots,g_{\nReceive,s}$ violates the degree constraints if there is no unique solution.
Thus, we declare a decoding failure if at least one polynomial has weighted degree $\deg_w$ at least $\nReceive-\tau$.

Compared to the matrix-based approach in Section~\ref{subsec:root-finding}, the detection of decoding failures is more efficient since computing the rank of a $d_I\times s$ matrix needs more operations than determining the degree of a linearized polynomial. 
Since we use all polynomials that are constructed after all iterations of ~\cite[Algorithm 1]{Xie2011General}, constructing the $s$ polynomials does not require more operations than constructing one polynomial.
Thus the number of required operations in $\Fqm$ is on the order of $\OCompl{s^2\nReceive(\nReceive-\tau)}$.

\begin{figure}
%
%
\begin{tikzpicture}

\begin{axis}[%
width=1.1\columnwidth,
height=0.85\columnwidth,
xmin=0,
xmax=40,
xlabel={Number of insertions $\insertions$},
compat=newest,
label style={anchor=near ticklabel, font=\footnotesize},
xmajorgrids,
major grid style={dashed},
ymin=0,
ymax=700000,
ylabel={Multiplications $\cdot 10^5$},
label style={inner sep=0}, 
ymajorgrids,
tick label style={font=\scriptsize},
legend style={at={(0.015,0.98)},anchor=north west,draw=black,fill=white,legend cell align=left,font=\footnotesize},
scaled ticks = false,
ytick={0,100000,200000,300000,400000,500000, 600000,700000},
yticklabels={0,1,2,3,4,5,6,7}
]
\addplot [
color=red,
only marks,
mark=o,
mark options={solid}
]
table[row sep=crcr]{
0 100800\\
1 102060\\
2 103320\\
3 104580\\
4 105840\\
5 113475\\
6 114810\\
7 116145\\
8 117480\\
9 118815\\
10 126900\\
11 128310\\
12 129720\\
13 131130\\
14 132540\\
15 141075\\
16 142560\\
17 144045\\
18 145530\\
19 147015\\
20 156000\\
21 157560\\
22 159120\\
23 160680\\
24 162240\\
25 171675\\
26 173310\\
27 174945\\
28 176580\\
29 178215\\
30 188100\\
31 189810\\
32 191520\\
33 193230\\
34 194940\\
35 205275\\
36 207060\\
37 208845\\
38 210630\\
39 212415\\
40 223200\\
};
\addlegendentry{$n_t\!=\!80,k\!=\!60$ Interpolation};

\addplot [
color=blue,
only marks,
mark=diamond,
mark options={solid}
]
table[row sep=crcr]{
0 200760\\
1 204336\\
2 207911\\
3 211484\\
4 215054\\
5 238595\\
6 242606\\
7 246616\\
8 250624\\
9 254629\\
10 280905\\
11 285376\\
12 289846\\
13 294314\\
14 298779\\
15 327940\\
16 332896\\
17 337851\\
18 342804\\
19 347754\\
20 379950\\
21 385416\\
22 390881\\
23 396344\\
24 401804\\
25 437185\\
26 443186\\
27 449186\\
28 455184\\
29 461179\\
30 499895\\
31 506456\\
32 513016\\
33 519574\\
34 526129\\
35 568330\\
36 575476\\
37 582621\\
38 589764\\
39 596904\\
40 642740\\
};
\addlegendentry{$n_t\!=\!80,k\!=\!60$ Gaussian Elimination};

\addplot [
color=red,
only marks,
mark=triangle,
mark options={solid,,rotate=270}
]
table[row sep=crcr]{
0 16320\\
1 16830\\
2 17340\\
3 20475\\
4 21060\\
5 21645\\
6 22230\\
7 22815\\
8 26400\\
9 27060\\
10 27720\\
11 28380\\
12 29040\\
13 33075\\
14 33810\\
15 34545\\
16 35280\\
17 36015\\
18 40500\\
19 41310\\
20 42120\\
21 42930\\
22 43740\\
23 48675\\
24 49560\\
25 50445\\
26 51330\\
27 52215\\
28 57600\\
29 58560\\
30 59520\\
31 60480\\
32 61440\\
33 67275\\
34 68310\\
35 69345\\
36 70380\\
37 71415\\
38 77700\\
39 78810\\
40 79920\\
};
\addlegendentry{$n_t\!=\!32,k\!=\!20$ Interpolation};

\addplot [
color=blue,
only marks,
mark=asterisk,
]
table[row sep=crcr]{
0 14736\\
1 15334\\
2 15929\\
3 20370\\
4 21156\\
5 21941\\
6 22724\\
7 23504\\
8 29180\\
9 30176\\
10 31171\\
11 32164\\
12 33154\\
13 40215\\
14 41446\\
15 42676\\
16 43904\\
17 45129\\
18 53725\\
19 55216\\
20 56706\\
21 58194\\
22 59679\\
23 69960\\
24 71736\\
25 73511\\
26 75284\\
27 77054\\
28 89170\\
29 91256\\
30 93341\\
31 95424\\
32 97504\\
33 111605\\
34 114026\\
35 116446\\
36 118864\\
37 121279\\
38 137515\\
39 140296\\
40 143076\\
};
\addlegendentry{$n_t\!=\!32,k\!=\!20$ Gaussian Elimination};

\end{axis}
\end{tikzpicture}%
 \caption{Multiplications vs. the number of insertions for $s=4$.}\label{fig:complexity}
 \vskip -5pt
\end{figure}
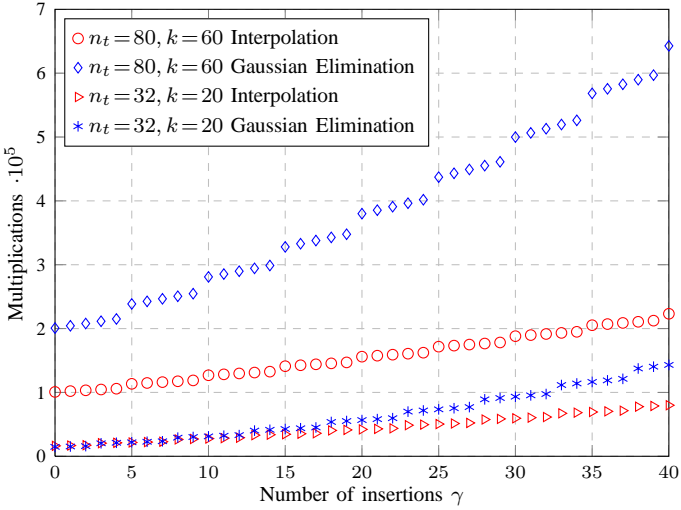

This interpretation of \cite{Xie2011General} can be directly applied to the decoding approach for interleaved Gabidulin codes in \cite{WachterzehZeh-InterpolationInterleavedGabidulin}.
Fig.~\ref{fig:complexity} shows the number of multiplications needed for efficient interpolation and compares this to the number of multiplications needed to solve \eqref{eq:intMatrix} by Gaussian elimination.
The figure shows that our interpolation-based decoding substantially reduces complexity for large $\nTransmit$.
%
\section{Efficient Root-Finding}\label{sec:efficientRootFinding}
In \cite{WachterzehZeh-InterpolationInterleavedGabidulin} it was shown that the root-finding system \eqref{eq:rootFindingSystem} can be solved recursively with at most $\OCompl{s^3k^2}$ operations in $\Fqm$.
In this section, we present an efficient root-finding algorithm for the probabilistic unique decoder that finds all roots $f^{(j)}(x),j\in\intervallincl{1}{s}$, of the $s$ polynomials $Q^{(j)}$ which are $y_j$-minimal. The complexity of this algorithm is $\OCompl{s^2k^2}$ operations in $\Fqm$.

\subsection{An Efficient Root-Finding Algorithm}
We know that the interpolation algorithm constructs $s$ polynomials $Q^{(j)}(x,y_1,\dots,y_s)$ that are $y_j$-minimal, i.e., their leading term in $y_j$ is minimal for $j\in\intervallincl{1}{s}$.
This property imposes a triangular structure on the root-finding system of equations which allows to solve it efficiently.
Thus, applying the efficient interpolation algorithm allows to reduce the complexity of the root-finding step.

Algorithm~\ref{alg:efficientRootFinding} recursively determines all unique roots $f^{(j)}(x),j\in\intervallincl{1}{s}$, if all $s$ polynomials $Q^{(j)}$ (constructed by the interpolation algorithm) have \kwDeg-weighted degree less that $\nReceive-\tau$. 
Algorithm~\ref{alg:efficientRootFinding} is based on the symbolic right division of univariate linearized polynomials and solves the root-finding system recursively.

\IncMargin{1em}
\begin{algorithm}
	\caption{Unique root-finding of $y_1,\dots,y_s$-minimal polynomials\newline $f^{(1)}(x),\dots,f^{(s)}(x)\gets$findRoots$(Q^{(1)}\!,\dots,Q^{(s)},k)$}\label{alg:efficientRootFinding}
	\SetKwInOut{Input}{input}
	\SetKwInOut{Output}{output}
	\Indm  
	\Input{$s$ linearized polynomials $Q^{(j)}(x,y_1,\dots,y_s)$ being $y_j$-minimal for $j\in\intervallincl{1}{s}$}
	\Output{$s$ unique linearized message polynomials $f^{(1)}(x),\dots,f^{(s)}(x)$, $\deg_q(f^{(j)}(x)) <k$}
	\BlankLine
	\Indp
	\For{$i\gets1$ \KwTo $k$}{
		\For{$j\gets 1$ \KwTo $s$}{
			$d \gets\deg_q(Q_0^{(j)}(x))$, $e \gets deg_q(Q^{(j)}_j(y_j))$\\
			\If{$d-e=k-i$}{
				$t^{(j)}_i(x) \gets \left(-\frac{LC(Q_0^{(j)}(x))}{LC(Q^{(j)}_j(y_j))}\right)^{[m-e]} \cdot x^{[d-e]}$\\	\label{line:monomial}
				$f^{(j)}(x) \gets f^{(j)}(x) + t^{(j)}_i(x)$ \\
				\For{$\ell\gets$ 1 \KwTo $s$}{
					$Q^{(\ell)}_0(x) \gets Q^{(\ell)}_0(x) + Q^{(\ell)}_{j}(x)\otimes t^{(j)}_i(x)$
				}
			}
		}
	}
\end{algorithm}\DecMargin{1em}

\begin{theorem}[Correctness of Algorithm~\ref{alg:efficientRootFinding}]
 \hspace{7em}
 Let $Q^{(j)}(x,y_1,\dots,y_s)$ be $y_j$-minimal with $\kwDeg$-weighted degree $deg_w(Q^{(j)}(x,y_1,\dots,y_s))<\nReceive-\tau$, $\forall j\in\intervallincl{1}{s}$.
 Then, for all $j\in\intervallincl{1}{s}$, Algorithm~\ref{alg:efficientRootFinding} determines the unique polynomials $f^{(j)}(x)=\sum_{i=0}^{k-1}f^{(j)}_{i}x^{[i]}$ such that
 \begin{equation*}
	P^{(j)}(x)\defeq Q^{(j)}_{0}(x)+Q^{(j)}_{1}(f^{(1)}(x))+\dots+Q^{(j)}_{s}(f^{(s)}(x))= 0.
 \end{equation*}
\end{theorem}
\begin{IEEEproof}
 Let $d=\deg_q(Q^{(j)}_0(x))$ and $e=\deg_q(Q^{(j)}_j(y_j))$.
 The polynomial $Q^{(j)}(x,y_1,\dots,y_s)$ is $y_j$-minimal for any $j\in\intervallincl{1}{s}$ and thus the $q$-degrees fulfill $d\leq e+k-1$ and $\deg_q(Q^{(j)}_{j'}(y_{j'}))<e$ for $j'>j$.
 This implies that the coefficients are $q^{(j)}_{j',e}=0$ for $j'>j$.
 In the first iteration $j=i=1$ we must solve $\sum_{\ell=1}^{s} q^{(1)}_{\ell,e}\cdot f^{(\ell)[e]}_{k-1}x^{[k+e-1]}=-q^{(1)}_{0,d}x^{[d]}$. 
 Since $q^{(1)}_{\ell,e}=0$ for $\ell\in\intervallincl{2}{s}$ the calculation reduces to $q^{(1)}_{1,e}\cdot f^{(1)[e]}_{k-1}x^{[k+e-1]}=-q^{(1)}_{0,d}x^{[d]}$.
 Hence, the monomial $t_1^{(1)}(x) = f^{(1)}_{k-1}x^{[k-1]}$ can uniquely be determined as
 \begin{equation}
  f^{(1)}_{k-1}x^{[k-1]}=\left(-\frac{q^{(1)}_{0,d}}{q^{(1)}_{1,e}}\right)^{[m-e]}x^{[d-e]},
 \end{equation}
 which corresponds to Line~\ref{line:monomial} in Algorithm~\ref{alg:efficientRootFinding} and it is easy to check that 
 \begin{align}\label{eq:killLeadingMonomial}
	&LT(Q^{(1)}_1(y_1)\otimes t_1^{(1)}(x)) = q^{(1)}_{1,e}\cdot(t_1^{(1)}(x))^{[e]} \nonumber \\
	=&q^{(1)}_{1,e}\cdot \left((-q^{(1)}_{0,d}/q^{(1)}_{1,e})^{[m-e]}x^{[d-e]}\right)^{[e]} = -q^{(1)}_{0,d}x^{[d]}.
 \end{align}
If $d-e<k-1$, then $f^{(1)}_{k-1}$ has to be zero to ensure $q^{(1)}_{1,e}\cdot f^{(1)[e]}_{k-1}x^{[k+e-1]}=-q^{(1)}_{0,d}$, i.e., no update on $f^{(1)}(x)$ is done.
 Now $Q^{(\ell)}_{0}(x), \forall\ell\in\intervallincl{1}{s}$ are updated with
 \begin{equation*}
	Q^{(\ell)'}_0(x) = Q^{(\ell)}_0(x) + Q^{(\ell)}_{1}(x)\otimes t^{(1)}_1(x).
 \end{equation*}
 which reduces the $q$-degree of $Q^{(1)}_0(x)$ by one since we enforced $\eqref{eq:killLeadingMonomial}$.
 
 In the next iteration $i=1,j=2$ we must solve $\sum_{\ell=1}^{s} q^{(2)}_{\ell,e}\cdot f^{(\ell)[e]}_{k-1}x^{[k+e-1]}=-q^{(2)}_{0,d}x^{[d]}$ which reduces to 
 \begin{align*}
	&\sum_{\ell=1}^{2} q^{(2)}_{\ell,e}\cdot f^{(\ell)[e]}_{k-1}x^{[k+e-1]}=-q^{(2)}_{0,d}x^{[d]} \\
		\Leftrightarrow \ &q^{(2)}_{2,e}\cdot f^{(2)[e]}_{k-1}x^{[k+e-1]} = -(q^{(2)}_{0,d}x^{[d]} + q^{(2)}_{1,e}\cdot f^{(1)[e]}_{k-1}x^{[k+e-1]}) \\
			\Leftrightarrow \ &q^{(2)}_{2,e}\cdot f^{(2)[e]}_{k-1}x^{[k+e-1]} = -q^{(2)'}_{0,d}x^{[d]}
 \end{align*}
 due to the $y_2$-minimality of $Q^{(2)}(x,y_1,\dots,y_s)$.
 Hence, we can directly calculate the coefficient $f^{(2)}_{k-1}$ using the updated polynomial $Q^{(2)'}_0(x)$ from the previous step.

 The algorithm computes the unique monomial $f^{(j)}_{k-i}x^{[k-i]}$ in each step $i,j$ such that
 \begin{align*}
	q^{(j)}_{j,e}\!\cdot\! f^{(j)[e]}_{k-i}x^{[k+e-i]}\! =\! -\!\Big(q^{(j)}_{0,d}x^{[d]} + \sum_{\ell=1}^{j-1}q^{(j)}_{\ell,e}\cdot f^{(\ell)[e]}_{k-i}x^{[k+e-i]}\Big)\!\!
 \end{align*}
 which is possible since the right hand side is computed in step $i, j-1$.
\end{IEEEproof}

\subsection{Complexity analysis}
Suppose we use a normal basis (which always exists).
The calculation of $q$-powers then corresponds to a cyclic shift over the base field and its complexity can be neglected.
Hence, multiplications dominate the complexity.

\begin{lemma}[Complexity of the Root-Finding Algorithm]
	Let  $Q^{(j)}(x,y_1,\dots,y_s)$ be $y_j$-minimal and let $\deg_w(Q^{(j)}(x,y_1,\dots,y_s))<\nReceive-\tau$,  $\forall j\in\intervallincl{1}{s}$.
		Then the $s$ unique message polynomials $f^{(j)}(x), j\in\intervallincl{1}{s}$, such that   
	\begin{equation*}
		P(x) = Q^{(j)}\left(x,f^{(1)}(x),\dots,f^{(s)}(x)\right) = 0
	\end{equation*}
	can be found with \OCompl{s^2k^2} operations in \Fqm.
\end{lemma}

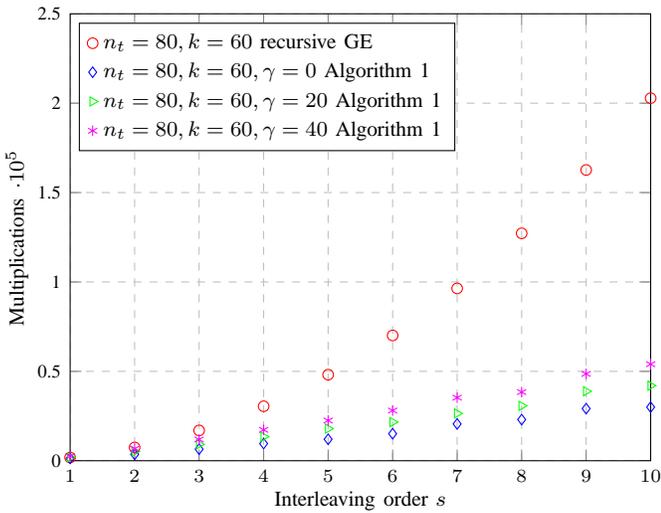
\begin{figure}[ht]
\begin{center}
%
%
%
\definecolor{mycolor1}{rgb}{1,0,1}%
\begin{tikzpicture}

\begin{axis}[%
width=1.05\columnwidth,
height=0.85\columnwidth,
xmin=1,
xmax=10,
xlabel={Interleaving order $s$},
compat=newest,
label style={anchor=near ticklabel, font=\footnotesize},
xmajorgrids,
major grid style={dashed},
ymin=0,
ymax=250000,
ylabel={Multiplications $\cdot 10^5$},
label style={inner sep=0}, 
ymajorgrids,
tick label style={font=\scriptsize},
legend style={at={(0.015,0.98)},anchor=north west,draw=black,fill=white,legend cell align=left,font=\footnotesize},
scaled ticks = false,
ytick={0,50000,100000,150000,200000,250000},
yticklabels={0,0.5,1,1.5,2,2.5},
xtick={1,2,3,4,5,6,7,8,9,10},
]
]
\addplot [
color=red,
only marks,
mark=o,
mark options={solid}
]
table[row sep=crcr]{
1 1830\\
2 7440\\
3 16950\\
4 30480\\
5 48150\\
6 70080\\
7 96390\\
8 127200\\
9 162630\\
10 202800\\
};
\addlegendentry{$n_t=80, k=60$ recursive GE};

\addplot [
color=blue,
only marks,
mark=diamond,
mark options={solid}
]
table[row sep=crcr]{
1 1320\\
2 3600\\
3 6480\\
4 9600\\
5 12000\\
6 15120\\
7 20580\\
8 23040\\
9 29160\\
10 30000\\
};
\addlegendentry{$n_t=80, k=60, \gamma = 0$ Algorithm \ref{alg:efficientRootFinding}};

\addplot [
color=green,
only marks,
mark=triangle,
mark options={solid,,rotate=270}
]
table[row sep=crcr]{
1 1920\\
2 5280\\
3 9180\\
4 13440\\
5 18000\\
6 21600\\
7 26460\\
8 30720\\
9 38880\\
10 42000\\
};
\addlegendentry{$n_t=80, k=60, \gamma = 20$ Algorithm \ref{alg:efficientRootFinding}};

\addplot [
color=mycolor1,
only marks,
mark=asterisk,
mark options={solid}
]
table[row sep=crcr]{
1 2520\\
2 6720\\
3 11880\\
4 17280\\
5 22500\\
6 28080\\
7 35280\\
8 38400\\
9 48600\\
10 54000\\
};
\addlegendentry{$n_t=80, k=60, \gamma = 40$ Algorithm \ref{alg:efficientRootFinding}};

\end{axis}
\end{tikzpicture}%
\caption{Number of multiplications for root-finding step for different interleaving orders. The complexity of the recursive GE is independent of $\insertions$.}\label{fig:complexityRF}
\end{center}
\end{figure}

\begin{IEEEproof}
Each step of Algorithm \ref{alg:efficientRootFinding} provides $s$ compositions of a (univariate) linearized polynomial of $q$-degree at most $\nReceive-\tau-k$ with one monomial.
If we consider the inversion as a multiplication we have $s(\nReceive-\tau-k+2)$ multiplications per step.
In total we need $ks^2(\nReceive-\tau-k+2)\leq k^2s^2$ multiplications in \Fqm.
\end{IEEEproof}
A comparison of this complexity to the recursive Gaussian elimination (recursive GE) from~\cite{WachterzehZeh-InterpolationInterleavedGabidulin} is illustrated in Figure~\ref{fig:complexityRF}.

Besides the reduction in computational cost, Algorithm~\ref{alg:efficientRootFinding} requires less memory than the recursive GE.
An upper bound on the memory requirement is given by the following lemma.
\begin{lemma}[Memory Requirement of Algorithm \ref{alg:efficientRootFinding}]
\hspace{5em} 
 The memory requirement of Algorithm \ref{alg:efficientRootFinding} is upper bounded by $s^2(\nReceive - \tau - k + 1) + s(\nReceive - \tau + k)$.
\end{lemma}
\begin{IEEEproof}
 The algorithm must store $s$ $(s+1)$-variate polynomials with each at most $((\nReceive - \tau) + s(\nReceive - \tau - k + 1)$ coefficients in $\Fqm$, as well as the $sk$ coefficients in $\Fqm$ of the $s$ message polynomials.
 Hence, in total we must store $s^2(\nReceive - \tau - k + 1) + s(\nReceive - \tau + k)$ elements in $\Fqm$.
\end{IEEEproof}

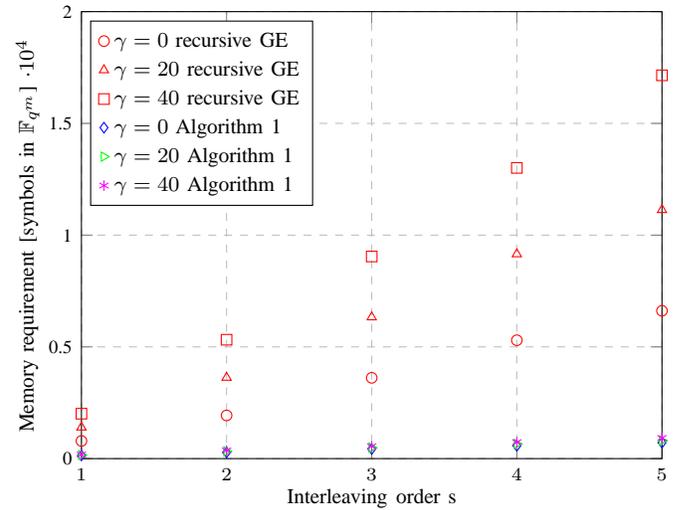
\begin{figure}[ht]
\begin{center}
%
%
%
\definecolor{mycolor1}{rgb}{1,0,1}%
\begin{tikzpicture}

\begin{axis}[%
width=1.05\columnwidth,
height=0.85\columnwidth,
xmin=1,
xmax=5,
xlabel={Interleaving order s},
xmajorgrids,
major grid style={dashed},
compat=newest,
label style={anchor=near ticklabel, font=\footnotesize},
ymin=0,
ymax=20000,
ylabel={Memory requirement [symbols in $\Fqm$] $\cdot 10^4$},
label style={inner sep=0}, 
ymajorgrids,
tick label style={font=\scriptsize},
xtick={1,2,3,4,5},
legend style={at={(0.015,0.98)},anchor=north west,draw=black,fill=white,legend cell align=left,font=\footnotesize},
scaled ticks = false,
ytick={0,5000,10000,15000,20000},
yticklabels={0,0.5,1,1.5,2}
]
\addplot [
color=red,
only marks,
mark=o,
mark options={solid}
]
table[row sep=crcr]{
1 790\\
2 1932\\
3 3615\\
4 5296\\
5 6615\\
};
\addlegendentry{$\insertions=0$ recursive GE};

\addplot [
color=red,
only marks,
mark=triangle,
mark options={solid}
]
table[row sep=crcr]{
1 1400\\
2 3626\\
3 6330\\
4 9152\\
5 11130\\
};
\addlegendentry{$\insertions=20$ recursive GE};

\addplot [
color=red,
only marks,
mark=square,
mark options={solid}
]
table[row sep=crcr]{
1 2010\\
2 5320\\
3 9045\\
4 13008\\
5 17150\\
};
\addlegendentry{$\insertions=40$ recursive GE};

\addplot [
color=blue,
only marks,
mark=diamond,
mark options={solid}
]
table[row sep=crcr]{
1 141\\
2 280\\
3 429\\
4 576\\
5 715\\
};
\addlegendentry{$\gamma = 0$ Algorithm \ref{alg:efficientRootFinding}};

\addplot [
color=green,
only marks,
mark=triangle,
mark options={solid,,rotate=270}
]
table[row sep=crcr]{
1 161\\
2 322\\
3 489\\
4 656\\
5 805\\
};
\addlegendentry{$\gamma = 20$ Algorithm \ref{alg:efficientRootFinding}};

\addplot [
color=mycolor1,
only marks,
mark=asterisk,
mark options={solid}
]
table[row sep=crcr]{
1 181\\
2 364\\
3 549\\
4 736\\
5 925\\
};
\addlegendentry{$\gamma = 40$ Algorithm \ref{alg:efficientRootFinding}};
\end{axis}
\end{tikzpicture}%
\caption{Memory requirements of the root-finding step for $\nTransmit=80, k=60$}\label{fig:memPlotRootFinding}
\vskip -12pt
\end{center}
\end{figure}

The memory requirements of Algorithm~\ref{alg:efficientRootFinding} and the recursive GE are illustrated in Figure~\ref{fig:memPlotRootFinding}.

\section{Efficient Decoding Algorithm}\label{sec:entire_eff_decoding}
We now summarize the efficient decoding procedure for list and probabilistic unique decoding of interleaved subspace and Gabidulin codes. 
The decoding procedure uses the efficient interpolation and root-finding algorithm for list decoding of KK codes and is summarized in Algorithm \ref{alg:listDec}.
\begin{algorithm}
	\caption{ListDecodeIntSub$(\vec{x}^{T}, \vec{r}^{(1)T}, \dots, \vec{r}^{(s)T})$}\label{alg:listDec}
	\DontPrintSemicolon
	\SetKwInOut{Input}{Input}\SetKwInOut{Output}{Output}
	\Input{a basis $(\vec{x}^{T}, \vec{r}^{(1)T}, \dots, \vec{r}^{(s)T})$ for the $\nReceive$-dimensional received subspace}
	\Output{A list $\mathcal{L}$ of tuples $(f^{(1)}(x),\dots,f^{(s)}(x))$}
	\BlankLine
	\textbf{Interpolation step:}\\
	$Q^{(1)}\!,\dots,\! Q^{(s)}\!\gets\!$ InterpolateBasis$(\vec{x}^{T}, \vec{r}^{(1)T}, \dots, \vec{r}^{(s)T})$ 
	\BlankLine
	\textbf{Root-finding step:}\\
	Pick $Q^{(1)}\!,\dots,Q^{(r)}$ with $\deg_w(Q^{(j)})<\nReceive\!-\!\tau,j\!\in\!\intervallincl{1}{r}$\\
	Set up the root-finding matrix $\Mat{Q}$ \eqref{eq:rootFindingMatrix} and $\vec{q}_0$ using $Q^{(1)}\!,\dots,Q^{(r)}$\\
	Determine all solutions of the root-finding system $\Mat{Q}\cdot\vec{f}=\vec{q}_0$, i.e. all roots of \eqref{eq:intPoly}\\
	\BlankLine
	\textbf{Output:} List $\mathcal{L}$ of all tuples $(f^{(1)}(x),\dots,f^{(s)}(x))$ which are a solution for the root-finding system \eqref{eq:rootFindingSystem}
\end{algorithm}
In order to efficiently decode interleaved Gabidulin codes of length~$n$, dimension $k$ and interleaving order $s$ as defined in \cite{Loidreau_Overbeck_Interleaved_2006,SidBoss_InterlGabCodes_ISIT2010,WachterzehZeh-InterpolationInterleavedGabidulin} we set $\nTransmit=\nReceive=n$.
Let $\vec{g}\!=\!\{g_0,\dots,g_{n-1}\} \!\subset\! \mathbb{F}_{q^{m}}$ with $n\leq m$ denote the linearly independent code locators of the interleaved Gabidulin code and denote by $\vec{y}^{(j)},j\in\intervallincl{1}{s}$ the elementary received words.
Then, Algorithm \ref{alg:listDec} called with $(\vec{g}^{T},\vec{y}^{(1)T}, \dots, \vec{y}^{(s)T})$ can decode errors of rank $t$ up to $t\leq\tau<\frac{s(n-k+1)}{s+1}$ (see \cite{WachterzehZeh-InterpolationInterleavedGabidulin}). 

The complete procedure for the probabilistic unique decoder for interleaved KK codes is given in Algorithm~\ref{alg:uniqueDec}.
To decode an interleaved Gabidulin code, this procedure must be called with $(\vec{g}^{T},\vec{y}^{(1)T}, \dots, \vec{y}^{(s)T})$.
As in~\cite{WachterzehZeh-InterpolationInterleavedGabidulin}, the decoder finds a unique solution with high probability for $t\leq\frac{s(n-k)}{s+1}$.

\begin{algorithm}
	\caption{UniqueDecodeIntSub$(\vec{x}^{T}, \vec{r}^{(1)T}, \dots, \vec{r}^{(s)T})$}\label{alg:uniqueDec}
	\DontPrintSemicolon
	\SetKwInOut{Input}{Input}\SetKwInOut{Output}{Output}
	\Input{a basis $(\vec{x}^{T}, \vec{r}^{(1)T}, \dots, \vec{r}^{(s)T})$ for the $\nReceive$-dimensional received subspace}
	\Output{$s$ linearized polynomials $f^{(1)}(x),\dots,f^{(s)}(x)$ or ``decoding failure''}
	\BlankLine
	\textbf{Interpolation step:}\\
	$Q^{(1)}\!,\dots,\! Q^{(s)}\!\gets\!$ InterpolateBasis$(\vec{x}^{T}, \vec{r}^{(1)T}, \dots, \vec{r}^{(s)T})$ 
	\BlankLine
	\textbf{Root-finding step:}\\
	\If{$\deg_w(Q^{(j)}(x,y_1\dots,y_s))<\nReceive-\tau,\forall j\in\intervallincl{1}{s}$}{
		$f^{(1)}(x),\dots,f^{(s)}(x)\gets$ findRoots$(Q^{(1)},\dots,Q^{(s)},k)$\;
		\textbf{Output:} $f^{(1)}(x),\dots,f^{(s)}(x)$
	}
	\Else{
		\textbf{Output:} decoding failure
	}
\end{algorithm}

%
\vspace{-4pt}
\section{Conclusion}\label{sec:conclusion}
An interpolation-based decoding scheme for interleaved subspace (KK) codes has been presented.
We have shown that interleaved subspace codes can be made more resilient against insertions as compared to the approach from~\cite{koetter_kschischang}. 
Our principle can be used as a (not necessarily polynomial-time) list decoder as well as a probabilistic unique decoder.
In both cases, the procedure consists of interpolating a set of multivariate linearized polynomials followed by a root-finding step.

The required linearized polynomials can be constructed efficiently with an adapted version of the general linearized K\"otter interpolation.
The procedure substantially reduces the computational complexity of the interpolation step.
Further, a computationally- and memory-efficient root-finding algorithm for the unique decoder was presented, which exploits the structure of the output of the interpolation algorithm. 

Both algorithms can also be used to accelerate interpolation-based decoding for interleaved Gabidulin codes from~\cite{WachterzehZeh-InterpolationInterleavedGabidulin}.

\bibliographystyle{IEEEtran}
\bibliography{interleaved_subspace}

\begin{thebibliography}{10}
\providecommand{\url}[1]{#1}
\csname url@samestyle\endcsname
\providecommand{\newblock}{\relax}
\providecommand{\bibinfo}[2]{#2}
\providecommand{\BIBentrySTDinterwordspacing}{\spaceskip=0pt\relax}
\providecommand{\BIBentryALTinterwordstretchfactor}{4}
\providecommand{\BIBentryALTinterwordspacing}{\spaceskip=\fontdimen2\font plus
\BIBentryALTinterwordstretchfactor\fontdimen3\font minus
  \fontdimen4\font\relax}
\providecommand{\BIBforeignlanguage}[2]{{%
\expandafter\ifx\csname l@#1\endcsname\relax
\typeout{** WARNING: IEEEtran.bst: No hyphenation pattern has been}%
\typeout{** loaded for the language `#1'. Using the pattern for}%
\typeout{** the default language instead.}%
\else
\language=\csname l@#1\endcsname
\fi
#2}}
\providecommand{\BIBdecl}{\relax}
\BIBdecl

\bibitem{koetter_kschischang}
R.~K\"otter and F.~R. Kschischang, ``{Coding for Errors and Erasures in Random
  Network Coding},'' \emph{IEEE Trans. Inf. Theory}, vol.~54, no.~8, pp.
  3579--3591, Jul. 2008.

\bibitem{silva_rank_metric_approach}
D.~Silva, F.~R. Kschischang, and R.~K{\"o}tter, ``{A Rank-Metric Approach to
  Error Control in Random Network Coding},'' \emph{IEEE Trans. Inf. Theory},
  vol.~54, no.~9, pp. 3951--3967, 2008.

\bibitem{KohnertKurz-LargeConstantDimensionCodes-2008}
A.~Kohnert and S.~Kurz, ``{Construction of Large Constant Dimension Codes with
  a Prescribed Minimum Distance},'' in \emph{Mathematical Methods in Computer
  Science}, ser. Lecture Notes in Computer Science.\hskip 1em plus 0.5em minus
  0.4em\relax Springer Berlin Heidelberg, 2008, vol. 5393, pp. 31--42.

\bibitem{Etzion2009ErrorCorrecting}
T.~Etzion and N.~Silberstein, ``{Error-Correcting Codes in Projective Spaces
  Via Rank-Metric Codes and Ferrers Diagrams},'' \emph{IEEE Trans. Inform.
  Theory}, vol.~55, no.~7, pp. 2909--2919, Jul. 2009.

\bibitem{Xia2009Johnson}
S.~Xia and F.~Fu, ``{Johnson Type Bounds on Constant Dimension Codes},''
  \emph{Des. Codes Cryptogr.}, vol.~50, no.~2, pp. 163--172, Feb. 2009.

\bibitem{Skachek2010Recursive}
V.~Skachek, ``{Recursive Code Construction for Random Networks},'' \emph{IEEE
  Trans. Inform. Theory}, vol.~56, no.~3, pp. 1378--1382, Mar. 2010.

\bibitem{Gadouleau2010ConstantRank}
M.~Gadouleau and Z.~Yan, ``{Constant-Rank Codes and Their Connection to
  Constant-Dimension Codes},'' \emph{IEEE Trans. Inform. Theory}, vol.~56,
  no.~7, pp. 3207--3216, Jul. 2010.

\bibitem{TrautmannManganielloRosenthal-OrbitCodes-2010}
A.~L. Trautmann, F.~Manganiello, and J.~Rosenthal, ``{Orbit Codes - A new
  Concept in the Area of Network Coding},'' in \emph{IEEE Information Theory
  Workshop 2019 (ITW 2012)}, Aug. 2010.

\bibitem{Etzion2011ErrorCorrecting}
T.~Etzion and A.~Vardy, ``{Error-Correcting Codes in Projective Space},''
  \emph{IEEE Trans. Inform. Theory}, vol.~57, no.~2, pp. 1165--1173, Feb. 2011.

\bibitem{Bachoc2012Bounds}
C.~Bachoc, F.~Vallentin, and A.~Passuello, ``{Bounds for Projective Codes from
  Semidefinite Programming},'' \emph{Adv. Math. Commun.}, vol.~7, no.~2, pp.
  127--145, May 2013.

\bibitem{Delsarte_1978}
P.~Delsarte, ``{Bilinear Forms over a Finite Field with Applications to Coding
  Theory},'' \emph{J. Combin. Theory}, vol.~25, no.~3, pp. 226--241, 1978.

\bibitem{Gabidulin_TheoryOfCodes_1985}
E.~M. Gabidulin, ``{Theory of Codes with Maximum Rank Distance},'' \emph{Probl.
  Inf. Transm.}, vol.~21, no.~1, pp. 3--16, 1985.

\bibitem{Roth_RankCodes_1991}
R.~M. Roth, ``{Maximum-Rank Array Codes and their Application to Crisscross
  Error Correction},'' \emph{IEEE Trans. Inf. Theory}, vol.~37, no.~2, pp.
  328--336, 1991.

\bibitem{Loidreau_Overbeck_Interleaved_2006}
P.~Loidreau and R.~Overbeck, ``{Decoding Rank Errors Beyond the Error
  Correcting Capability},'' in \emph{Int. Workshop Alg. Combin. Coding Theory
  (ACCT)}, Sep. 2006, pp. 186--190.

\bibitem{SidBoss_InterlGabCodes_ISIT2010}
V.~R. Sidorenko and M.~Bossert, ``{Decoding Interleaved Gabidulin Codes and
  Multisequence Linearized Shift-Register Synthesis},'' in \emph{IEEE Int.
  Symp. Inf. Theory (ISIT)}, Jun. 2010, pp. 1148--1152.

\bibitem{WachterzehZeh-InterpolationInterleavedGabidulin}
A.~Wachter-Zeh and A.~Zeh, ``{List and Unique Error-Erasure Decoding of
  Interleaved Gabidulin Codes with Interpolation Techniques},'' \emph{accepted
  for Des. Codes Cryptogr.}, 2014.

\bibitem{Wachterzeh_BoundsListDecodingRankMetric_IEEE-IT_2013}
A.~{Wachter-Zeh}, ``{Bounds on List Decoding of Rank-Metric Codes},''
  \emph{IEEE Trans. Inform. Theory}, vol.~59, no.~11, pp. 7268--7277, Nov.
  2013.

\bibitem{Mahdavifar2010Algebraic}
H.~Mahdavifar and A.~Vardy, ``{Algebraic List-Decoding on the Operator
  Channel},'' in \emph{IEEE Int. Symp. Inf. Theory}, Jun. 2010, pp. 1193--1197.

\bibitem{Mahdavifar2012Listdecoding}
------, ``{List-Decoding of Subspace Codes and Rank-Metric Codes up to
  Singleton Bound},'' in \emph{IEEE Int. Symp. Inf. Theory}, Jul. 2012, pp.
  1488--1492.

\bibitem{GuruswamiXing-ListDecodingRSAGGabidulinSubcodes_2012}
V.~Guruswami and C.~Xing, ``{List Decoding Reed--Solomon, Algebraic-Geometric,
  and Gabidulin Subcodes up to the Singleton Bound},'' \emph{Electronic Colloq.
  Comp. Complexity}, vol.~19, no. 146, 2012.

\bibitem{TrautmannSilbersteinRosenthal-ListDecodingLiftedGabidulinCodes}
A.-L. Trautmann, N.~Silberstein, and J.~Rosenthal, ``{List Decoding of Lifted
  Gabidulin Codes via the Pl\"ucker Embedding},'' in \emph{Int. Workshop Coding
  Cryptogr. (WCC)}, Apr. 2013.

\bibitem{Xie2011General}
H.~Xie, Z.~Yan, and B.~W. Suter, ``{General Linearized Polynomial Interpolation
  and Its Applications},'' in \emph{IEEE Int. Symp. Network Coding (Netcod)},
  Jul. 2011, pp. 1--4.

\bibitem{Ore_OnASpecialClassOfPolynomials_1933}
{\O}.~Ore, ``{On a Special Class of Polynomials},'' \emph{Trans. Amer. Math.
  Soc.}, vol.~35, pp. 559--584, 1933.

\bibitem{Lidl-Niederreiter:FF1996}
R.~Lidl and H.~Niederreiter, \emph{{Finite Fields}}, ser. Encyclopedia of
  Mathematics and its Applications.\hskip 1em plus 0.5em minus 0.4em\relax
  Cambridge University Press, Oct. 1996.

\end{thebibliography}
\end{document}